\documentclass[12pt]{article}
\input{epsf.tex}
\begin{document}
\begin{center}
{\bf 
PARAMETERS OF THE BEST APPROXIMATION FOR DISTRIBUTION OF THE REDUCED NEUTRON
WIDTHS. THE MOST PROBABLE DENSITY OF NEUTRON RESONANCES IN ACTINIDES}
\end{center}
\begin{center}
{\bf A.M. Sukhovoj, V.A. Khitrov}
\end{center}
\begin{center}
Joint Institute for Nuclear Research, Dubna, Russia, 141980
\end{center} 
\begin{abstract}
   In the frameworks of hypothesis of practical constancy of the neutron
resonance number in small fixed intervals $\Delta E$  of neutron energy,  their most probable
value was determined for nucleus mass region  $231 \leq  A \leq 243$  from approximation
of the reduced neutron widths by
superposition of two or  four independent distributions.
This was done under assumption that a set of the measured neutron amplitudes
can correspond to one or to superposition of some normal distributions with
non-zero average and dispersion differing from $<\Gamma_n^0>$.

The main result of the analysis: the mean $D$ and $S$ values can be
determined only with unknown systematical uncertainty whose magnitude 
is determined by unknown precision  of the
Porter-Thomas hypothesis correspondence to concrete experimental
sets of resonances and unknown experimental mean $<\Gamma_n^0>$ widths.
\end{abstract}
\section{Introduction}\hspace*{16pt}

The density of neutron resonances $\rho_{\lambda}=D_{\lambda}^{-1}$ is one of the main results of analysis
of the data of all experiments performed using the neutron time-of-flight
method. It is the basis point for any experiments where nucleus level density
is derived from the spectra of gamma-quanta or evaporation nucleons.
High precision in determination of $D_{\lambda}^{-1}$ is stipulated by excellent resolution
of corresponding method, but it can be realized only by careful accounting or
correction of all systematical errors of experiment.

The most serious and not removable from them is ``omission" of resonances whose
reduced neutron width $\Gamma_n^0$ ($\Gamma_n^1$...) is less than the sensitivity threshold of experiment.
In principle, determination of the most probable value of $D_{\lambda}^{-1}$ in this situation is possible
only by means of the most precise approximation of distribution of  $\Gamma_n^0$ in all
region of their values and extrapolation of the obtained function into the
region below threshold. Of course, precision of this procedure is determined
by degree of correspondence of theoretical notions about distribution $\Gamma_n^0$ to
experiment.

According to theoretical analysis \cite{PEPAN-1972}, the value  $\Gamma_n^0$ in nuclei of
intermediate and large mass is determined by few-quasi-particle components of
wave function whose square contribution in its normalization is estimated by
value of about $10^{-6}$-$10^{-9}$. Their small and chaotic values for different
resonances are determined by strong fragmentation \cite{MalSol} of low-lying
one- and two-quasi-particle states of a nucleus. There is the first necessary
condition for description of fluctuations of $\Gamma_n^0$ by the Porter-Thomas distribution \cite{PT}.

Another condition is that the mathematics expectation of mean value of amplitude $A=\sqrt{\Gamma_n^0}$
must be equal to zero, and its dispersion -- to mean $<\Gamma_n^0>$. Both conditions:
\begin{eqnarray}
M(A)=0, \nonumber\\
D(A)=<\Gamma_n^0>
\end{eqnarray}
are not tested in modern analysis of the experimental $\Gamma_n^0$  values
\cite{2008De20}.
I.e., applicability of the Porter-Thomas distribution is postulated but is
not proved. The experimental width distribution is not tested also for possibility
of existence of superposition of several distributions with different values
$M(A)$ and  $D(A)$.
	Approximation  \cite{Prep196, PEPAN-2006}
 of level density derived from the two-step
cascade intensities  shows that the structure of any nucleus changes cyclically
as increasing excitation energy.
This fact is determined so far as at present there is the only methodically
model-free method for determination of $\rho$ -- \cite{PEPAN-2005}.
	This occurs, at least, due to excitation of nucleus states with 
increasing number of quasi-particles and, probably, due to variation of number
and type of phonons. Fragmentation of these complicating nucleus states
inevitably changes coefficients of wave functions of neutron resonances
(as it follows from basis notions of quasi-particle-phonon model of nucleus).
As a result, it is possible violation of the Porter-Thomas distribution in
existing today interpretation (1).

\section{Data of analysis}\hspace*{16pt}

The method for analysis of the data on  $\Gamma_n^0$ accounting for these factors is
described in \cite{AMSa}, concrete results of the best fitting of the experimental data
for actinides are given in \cite{AMSb}.
Cumulative sum of  $\Gamma_n^0$ in suggested there analysis is approximated by one or
several distributions of the variables:
\begin{equation}
X=((A-b)/\sigma)^2
\end{equation} 
with the initial values of fitted parameters
\begin{eqnarray}
b=M(A) \neq 0, \nonumber\\
\sigma^2=D(A) \neq <\Gamma_n^0>.
\end{eqnarray}

Parameters of the best approximation of distribution of the experimental
values of $\Gamma_n^0$ in actinides for variants of their one ($K=1$)
or, maximum, four ($K=4$) distributions
with different $M$  and  $D$ are compared in \cite{AMSb}  between themselves or with
approximation of the distorted by given registration threshold pure model
random values. This analysis brings to the conclusion that at present it is
inadmissibly to exclude a possibility of existence of superposition of several
differing by parameters $b$ and  $\sigma$ width distributions in every nucleus.
Although unambiguous conclusion about its presence cannot be made on the basis
of the modern experimental data on the resonance neutron widths.
Therefore, the mean spacing  between resonances in actinides is determined
below in different ($K=2$ and $K=4$) variants.
	The suggested in \cite{AMSb} possibility to estimate the most
probable number of omitted
resonances in any experiment calls no doubts if only the functional dependence
of their part $\Delta S_{th}$  from the total number  $S$   was set on the
grounds of some data for concrete intervals of resonance energies.
Then the parameters of unknown distributions are determined by equation:
\begin{equation}
\chi^2=(S-(\psi(A,b,\sigma)-\Delta \psi_{th}))^2
\end{equation}

Here $\psi(A,b,\sigma)=\int{X*\Gamma(X)dX}$  for gamma-function
$\Gamma$ with variable $X$.
 The value $\Delta \psi_{th}$  is determined only by difference  $N_{t}-N_{exp}$ for the varied expected resonance $N_{t}$
number  in interval $\delta E$  and the obtained experimentally $N_{exp}$.
The number of these intervals practically was varied from 5 to 20 in dependence on
quantity of experimental values of widths.
Moreover, negative values $N_{t}-N_{exp}$  in all cases were changed by
zero.

The calculated and experimental cumulative sums in this equation have
differing values of variables: function $S$ was obtained under assumption
that the unknown mean value of neutron width corresponds to $\sum N_{exp}$,
but the mean neutron width   for the calculated value is determined by
sum $\sum N_{t}$.
Therefore, calculation of $\chi^2$ is carried out after corresponding
change in variable $X$ for difference $\psi-\Delta \psi_{th}$.

The serious enough problem is setting of dispersion of cumulative sum for
arbitrary value $X$. Methodically this problem has simple solution: there
are generated large sets of cumulative sums of squares of normally
distributed numbers with given $b$ and $\sigma$ values for each  ``partial"
function number $K$ and for them by means of usual relations of mathematical statistics
is determined function $\sigma=f(X)$ for each magnitude of variable $X$.
But, in practice, this procedure requires unacceptable expenditures of
computer time. That is why, 
possible change of the $\chi^2$ value for different densities of neutron
resonances for realistic values of dispersions of cumulative sums was
performed only for $^{232}$Th, $^{233,235}$U and $^{239}$Pu (only in
approach of validity of the Porter-Thomas distribution). 

The difference of principle between the results of this approximation
and the data given below was not revealed.

Function (4) has not real minimum and in this variant of analysis of
distributions of reduced neutron widths. Comparison between the calculated
and experimental cumulative sums shows that some small difference of
$\chi^2$ for tested $N_t$ values is mainly caused by strong fluctuations
of cumulative sums in region of the largest $X$ values.

Naturally, function $\Delta \psi$ can take into account and other factors
distorting experimental distribution of widths.
This accounting can be performed in frameworks of both some model approaches
and concrete experimental data. Of course, function $\Delta \psi$ cannot be
set unambiguously
for the majority of factors which distort the neutron widths distributions.

       The desired $D=\sum\delta E/\sum N_{t}$ value corresponds to minimum of
$\chi^2$ for varied values $D$.
Fluctuations of different strength in the found function $\chi^2=f(D)$
are connected with ambiguity of the best fit in the region of the large
$X$ values or change in parameters for elements of the tested
superposition at $K>1$. In particular, at change of $D$ in case $K=2$,
for example, the smaller values of $\chi^2$ can be really realized not for two,
but in fact -- three distributions: sum of widths distributions for both
spin values of resonances and additional distribution of widths
corresponding to the largest values of $\Gamma_n^0$ and parameter $b>>1$.

\begin{figure}\begin{center}
\vspace{4cm}
\leavevmode
\epsfxsize=14cm

\epsfbox{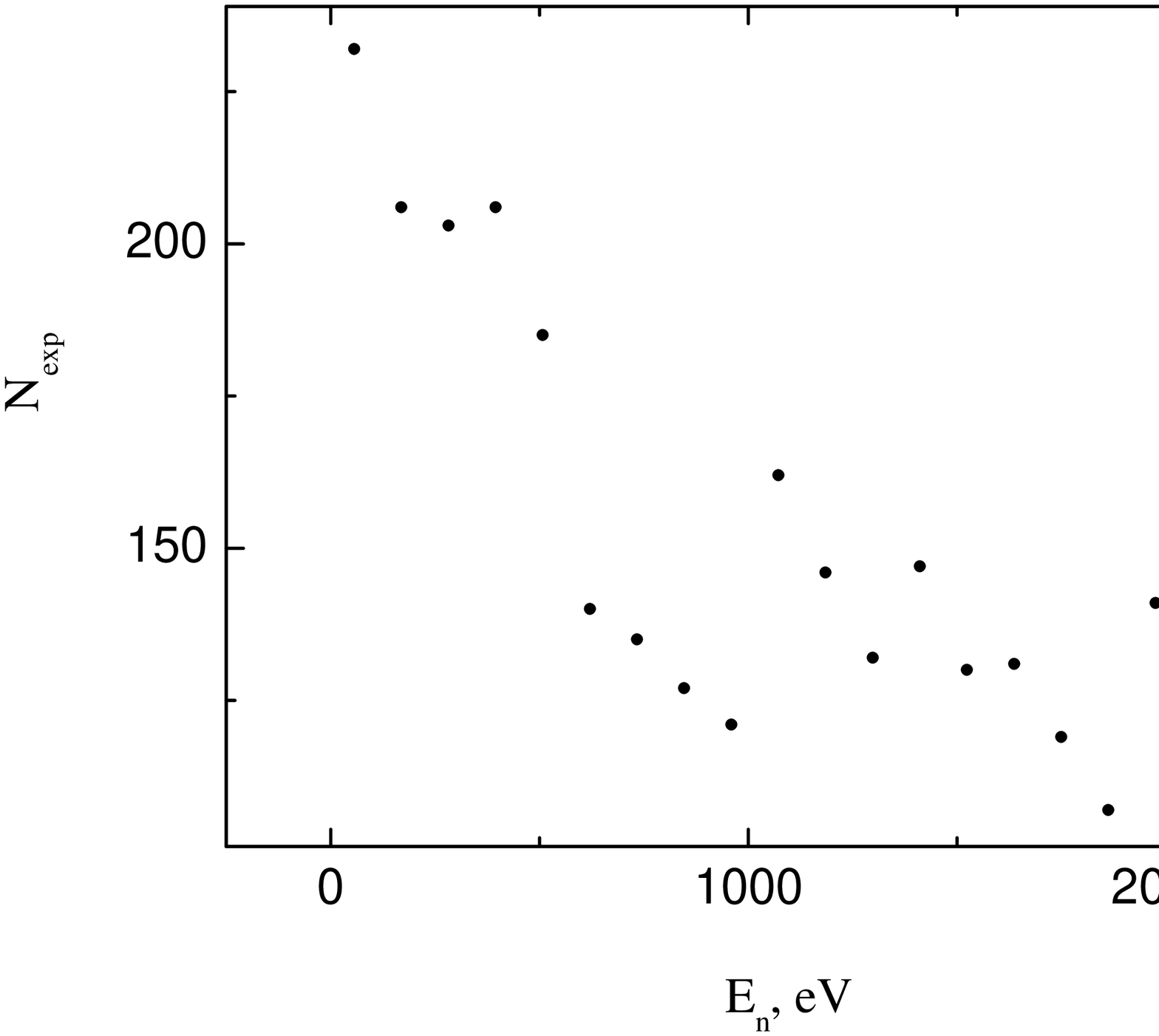} 

\end{center}
\vspace{-5cm}

Fig. 1. Experimental number of resonances in interval
$\delta E=113$ eV for $^{235}$U.
\end{figure}

       The example of concrete dependence of $N_{exp}$   is shown in Fig. 1.
The parameter of analysis (4) for this nucleus was tested for interval
$110 \leq N_t \leq 5000$.

 Comparison of experimental
cumulative sum of widths in $^{235}$U corresponding to different
expected density of neutron resonances for $D=0.1$ and $D=0.7$ eV
with the best approximation by expression (4) is presented in Fig. 2.

\begin{figure}

\leavevmode
\epsfxsize=14cm
\epsfbox{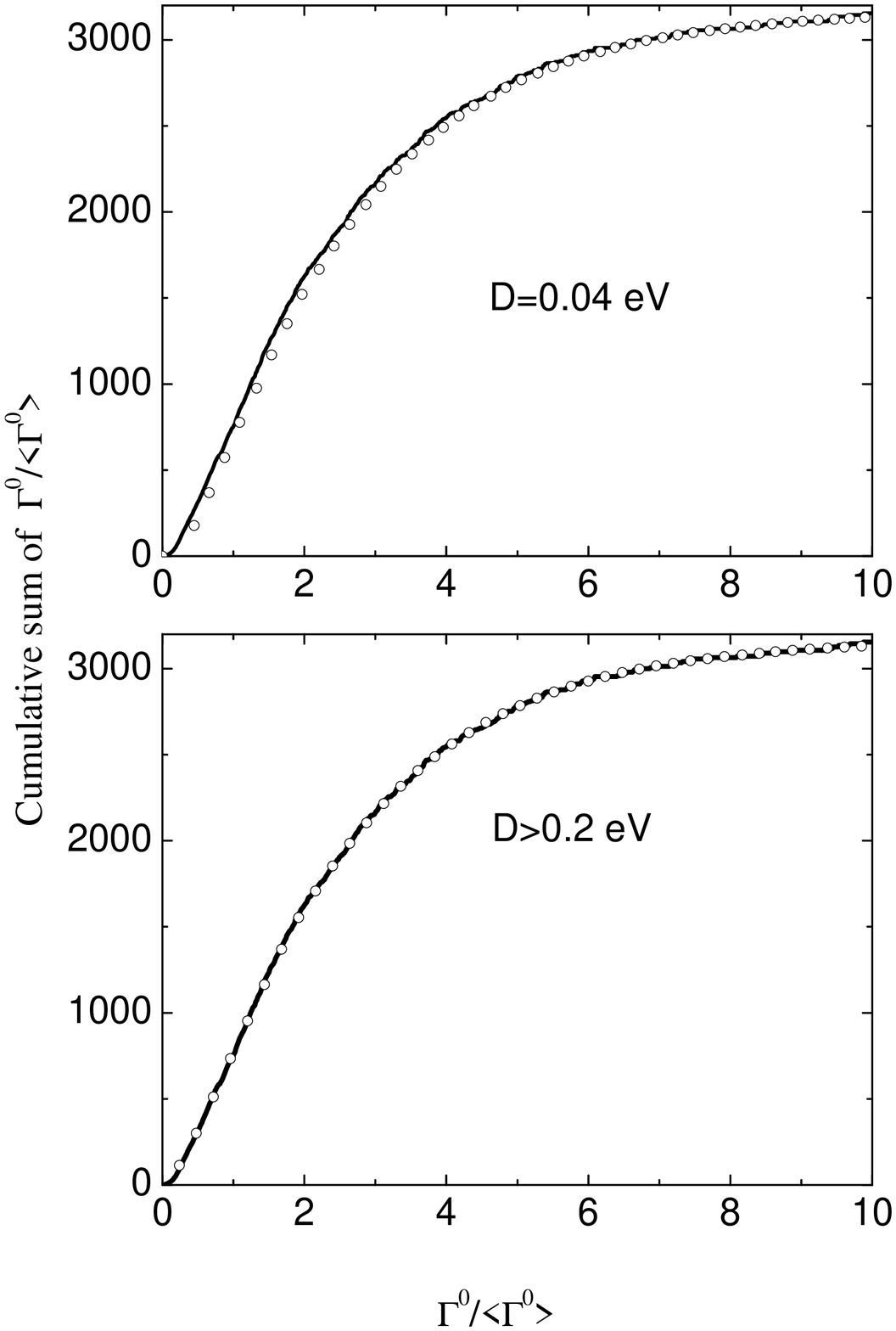} 

Fig. 2. Typical forms of the best approximations of cumulative sums
for the experimental data on the reduced neutron widths.
As an example, there are presented the data for $K=4$ $^{235}$U
in region of strong increase of $\chi^2$ and region of its practically
constant value.

\end{figure}

 In this nucleus, as in all investigated here nuclei,
is observed typical result: for $D\geq 0.1-0.2$ (odd) or $D \geq 1-2$
eV (even-even targets) is achieved the best and practically the same
degree of correspondence of the experiment and model approximation.
Any values of $b$ and $\sigma$ at noticeably smaller values $D$ cannot
give small $\chi^2$ by use of superposition of both
two and four different distributions. However, the values $\chi^2$ for
$K=2$, respectively, increase with respect to $K=4$. Sometimes --
very essentially. 

Besides, it should be taken into account that the practical search
of parameters $b$ and $\sigma$, which guarantees minimum of $\chi^2$
in the used method of approximation cannot secure the best approximation
of the experimental data in arbitrary variant of calculation.
Only the repeated variation of initial values and ways of random
processes can provide the sufficient for practical applications
precision of determination of the lowest possible $\chi^2$ value.

The obtained by us distributions  $\chi^2=f(D)$ for different variants of approximation of the experimental
data for nuclei from the mass region  $231 \leq A \leq 243$ are given in
figures 3-4. 

Estimated values of widths were taken from library ENDF/B-VII \cite{IAEA}.
In order to compensate ``omitted" resonances in $^{232}$Th  and $^{238}$U,
the authors of
the neutron resonance evaluation included for these nuclei in the library data
the resonances whose random widths are less than registration threshold.
Naturally, presented here analysis of such mixture may gives somewhat distorted information on
density of neutron resonances and is added below, most probably, for
demonstration of potential of the suggested method.
	Results of fitting of the  $D$ value, as it is seen from the data
presented
in figures 3-4 for each nucleus, depend on model notions.
In practice, one can conclude that:

(a) the analysis gives wide spectrum of possible $D$ values corresponding to
either practically constant $\chi^2$ value or -- weakly fluctuating function
of this parameter;

(b) weak local minima of $\chi^2$ are caused by bad stipulation of
approximation process for variant $K > 1$ distributions.

In both cases the number of fitted parameters is many times less than the number
of analyzed resonances. Therefore, the data for four distributions can be
adopted as the most probable ones.

\begin{figure}

\vspace{4.5cm}
\leavevmode
\epsfxsize=19cm
\hspace{-1.5cm}
\epsfbox{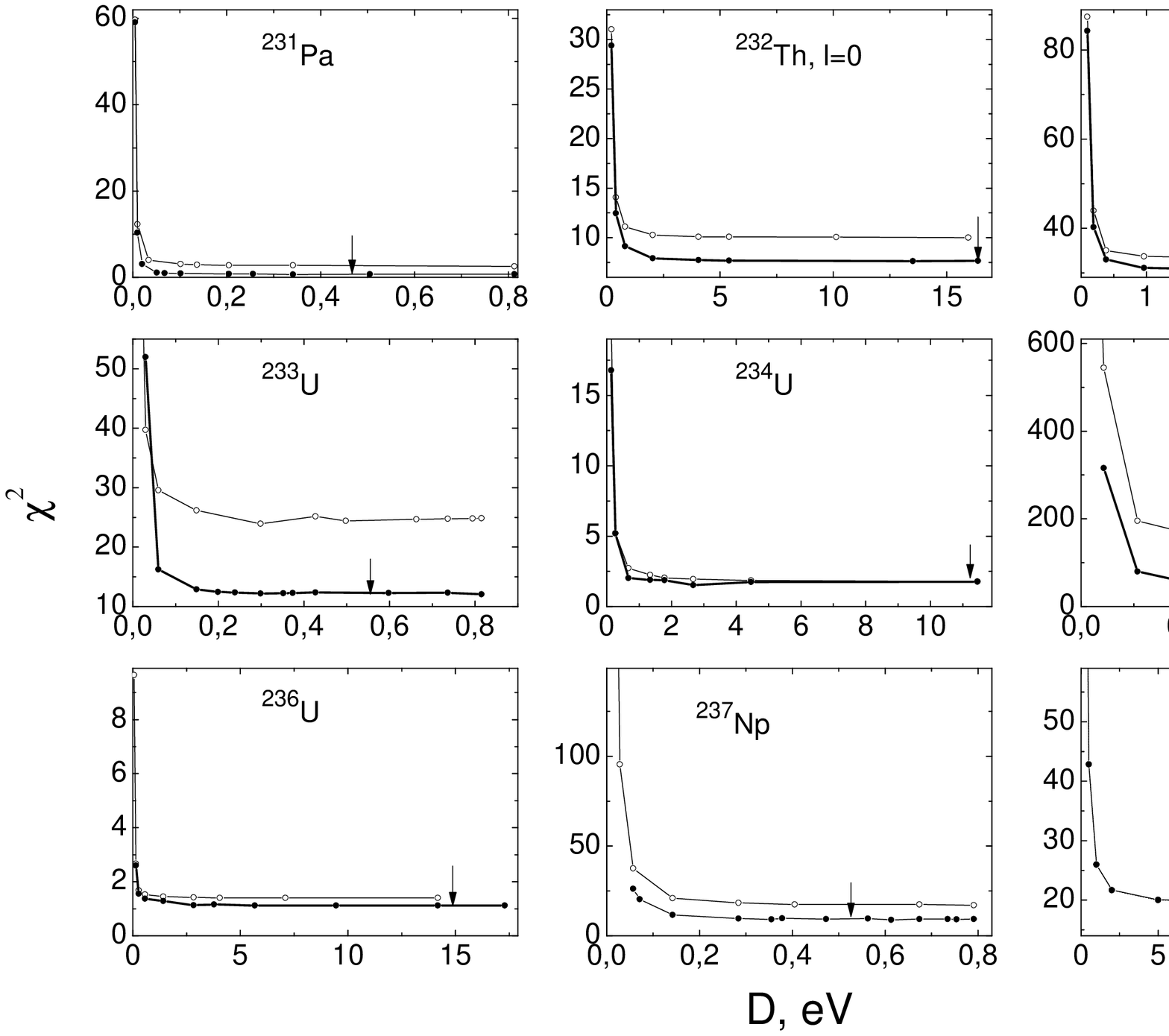} 
\vspace{-6cm}

Fig. 3. The $\chi^2$ value for the tested $D$ parameter for the nuclei with
mass $231 \leq A \leq 238$. The experimental sum of widths is approximated
by two (open) or four (full circles) distributions with corresponding magnitudes of variable (2).
The arrows correspond to $D$ values from [11] or [10].

\end{figure}

\newpage

\begin{figure}

\vspace{4.5cm}
\leavevmode
\epsfxsize=19cm
\hspace{-1.5cm}
\epsfbox{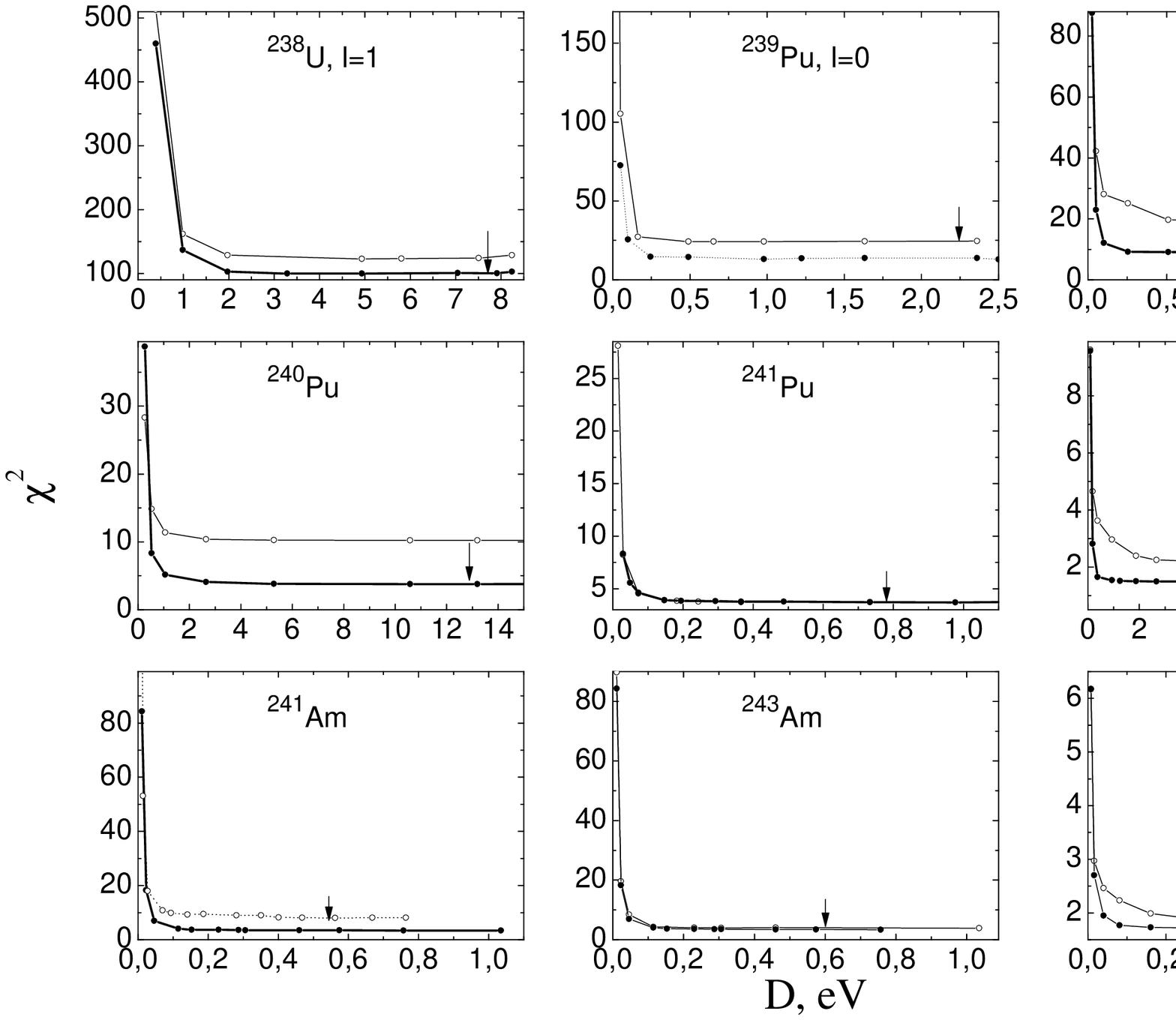} 
\vspace{-7cm}

Fig. 4. The same, as in Fig. 3, for nuclei with mass $238 \leq A \leq 243$.
\end{figure}

\section{Some items of fundamental problem of determination neutron
resonance widths distribution parameters}\hspace*{16pt}

The most important result obtained in frameworks of described analysis
of the experimental data on values of $\Gamma^0_n$ or $\Gamma^1_n$ --
the mean $\Gamma_n$ and $\rho$ values are at present determined with on
principle unknown systematical error. Really this result is expected:
parameters of any process under study cannot be found even from
mathematically strong extrapolation (or interpolation) of corresponding
data (in given case -- for the studied regions of nucleus excitation energy).
The unexpected point was the found here possibility that the mean value
of widths can be much less than registration threshold of experiment
.

In original paper \cite{PT} is stated without any proof that:
``As a consequence of experimental limitations, levels with small widths
may escape
detection, and also there may be only few of them...". Authors bring as
an example for $X=0.01$ the estimation of deviation in 9\% between the average
over measured widths and the expected one's average over the total distribution.
These statements are quite true in case of small part of widths which are
less than the threshold value. And they are absolutely mistaken --
in case when the main part of neutron widths lie below registration
threshold of experiment. Belonging of the tested set to one of these
extreme (as and intermediate) cases is determined by value of
$<\Gamma^0_n>$. In turn, it can be obtained only on the basis of necessary
amount of additional experimental information. 

Accordingly, all the published estimates of density of neutron resonances
contain unknown systematical error. In the best case it is enough
(for practical aims, for example) small; in the worse -- changes the values
of $\Gamma_n$
 and  $\rho$ by many times.
 The errors of parameters under consideration anticorrelate with each other.
Accordingly, at calculation of, for example, averaged  neutron-interaction
cross-sections, their uncertainties can be negligibly small  even for large
$\delta\Gamma_n$ and $\delta \rho$. However, for understanding of occurring
in nucleus processes of interaction and transition between Bose and Fermi
systems and determining them properties of nuclear matter, the achieved
precision for determination of level density can be insufficient.

Presentation of experimental data in form of cumulative sums of $\Gamma^0_n$
chosen for analysis has the lowest dependence on error of determination of
$<\Gamma^0_n>$. 
Therefore, the result obtained here could not be determined earlier
in the simplest analysis methods of distributions of $<\Gamma^0_n>$.

 As a consequence, {\bf any method for determination of mean parameters
of neutron widths distributions can give only some their probabilistic
values}.

\section{ Conclusion}\hspace*{16pt}

The main result of the neutron widths distribution analysis:
the $D$ and $S$ values can be
determined only with unknown systematical uncertainty whose magnitude 
is determined now by unknown precision  of the
Porter-Thomas distribution correspondence to concrete experimental
mean $<\Gamma_n^0>$ widths.

1. The suggested approximation of the total set of all the existing data on
widths of neutron resonances does not allow one to find unambiguously determined
absolute minimum of $\chi^2$ for the unique value of $D$.

2. The use for this aim of superposition of several distributions with the
different average and dispersion allows one to obtain the lowest value of $\chi^2$,
first of all, for the experimental data with number of widths exceeding $\sim 100$.

3. The analysis performed shows that the probability of correspondence of the
distribution $\Gamma^0_n$ to the simple functional dependence in nuclei of different mass is
less than that for the set of noticeably differing functions.
Therefore, any quantitative determination of parameters of their distribution
should be performed by comparison of two or more different model notions in
maximum set of nuclei.
 
4. The obtaining of the more unambiguous conclusions with respect to the
problem considered here requires very significant increase of sets of
resonances with the experimentally determined values
$\Gamma^0_n$ ($\Gamma^1_n$) at their
correspondingly decreased distortions.

5. The increase of precision for determination of the mean parameters of
the neutron width distributions requires, most probably, considerable
making more precise of model notions \cite{PT}. First of all of degree of influence
of structure of the nuclear excited levels on level density and probability of
emission of the nuclear reaction products in wide diapason of their energy.
In particular -- in region of neutron resonances. 

6. Selection of neutron resonances by orbital momentum must be performed in
common -- by minimum sum value of $\chi^2$ for obtained distributions
with $l=0$ and $l=1$, for example.
  

\end{document}